\input harvmac
\input psfig
\newcount\figno
\figno=0
\def\fig#1#2#3{
\par\begingroup\parindent=0pt\leftskip=1cm\rightskip=1cm\parindent=0pt
\global\advance\figno by 1
\midinsert
\epsfxsize=#3
\centerline{\epsfbox{#2}}
\vskip 12pt
{\bf Fig. \the\figno:} #1\par
\endinsert\endgroup\par
}
\def\figlabel#1{\xdef#1{\the\figno}}
\def\encadremath#1{\vbox{\hrule\hbox{\vrule\kern8pt\vbox{\kern8pt
\hbox{$\displaystyle #1$}\kern8pt}
\kern8pt\vrule}\hrule}}
\def\underarrow#1{\vbox{\ialign{##\crcr$\hfil\displaystyle
 {#1}\hfil$\crcr\noalign{\kern1pt\nointerlineskip}$\longrightarrow$\crcr}}}
%
\overfullrule=0pt

%
\def\tilde{\widetilde}
\def\bar{\overline}
\def\hat{\widehat}
\def\Z{{\bf Z}}

\def\S{{\bf S}}
\def\R{{\bf R}}

\font\zfont = cmss10 

\def\bigone{\hbox{1\kern -.23em {\rm l}}}
\def\ZZ{\hbox{\zfont Z\kern-.4emZ}}

\Title{hep-th/0307041} {\vbox{\centerline{$SL(2,{Z})$ Action On
Three-Dimensional    }
\bigskip
\centerline{Conformal Field Theories With Abelian Symmetry   }}}
\smallskip
\centerline{Edward Witten}
\smallskip
\centerline{\it Institute For Advanced Study, Princeton NJ 08540 USA}


\medskip

\noindent
On the space of three-dimensional conformal field theories with
$U(1)$ symmetry and a chosen coupling to a background gauge field,
there is a natural action of the group $SL(2,{\bf Z})$.  The
generator $S$ of $SL(2,{\bf Z})$ acts by letting the background
gauge field become dynamical, an operation considered recently by
Kapustin and Strassler in explaining three-dimensional mirror
symmetry. The other generator $T$ acts by shifting the
Chern-Simons coupling of the background field. This $SL(2,{\bf
Z})$ action in three dimensions is related by the AdS/CFT
correspondence to $SL(2,{\bf Z})$ duality of low energy $U(1)$
gauge fields in four dimensions.

\Date{June, 2003}
\newsec{Introduction}

In this paper, we will consider objects of the following kind:
conformal field theories\foot{In fact, we can extend the
definitions beyond conformal field theories to possibly massive
theories obtained by relevant deformations of conformal field
theories. However, the transformations of conformal fixed points
are a basic case, so we will phrase our discussion in terms of
conformal field theories.} in three dimensions that have a $U(1)$
symmetry, with an associated conserved current $J$. The goal of
the discussion is to show that there is a natural action of
$SL(2,{\bf Z})$ mapping such theories to other theories of the
same type.  This is not a duality group in the usual sense; an
$SL(2,{\bf Z})$ transformation does not in general map a theory to
an equivalent one or even to one that is on the same component of
the moduli space of conformal theories with $U(1)$ symmetry. It
simply maps a conformal field theory with $U(1)$ symmetry to
another, generally inequivalent conformal field theory with $U(1)$
symmetry.

Roughly speaking, we will give three approaches to understanding
the role of $SL(2,{\bf Z})$.  In section 3, after explaining how
the generators of $SL(2,{\bf Z})$ act, we perform a short formal
computation using properties of Chern-Simons theories to show that
the relevant operations  do in fact generate an action of
$SL(2,{\bf Z})$.  In section 4, for theories in which the current
two-point function is nearly Gaussian, we show directly the
$SL(2,{\bf Z})$ action on this two-point function.  Finally, in
section 5, we explain the origin of the $SL(2,{\bf Z})$ action
from the point of view of the AdS/CFT correspondence.

\nref\bd{C. Burgess and B. P. Dolan, ``Particle Vortex Duality And
The Modular Group: Applications To The Quantum Hall Effect And
Other 2-D Systems,'' hep-th/0010246.}%
\nref\wilczek{A. Shapere and F. Wilczek, ``Self-Dual Models With
Theta Terms,'' Nucl. Phys. {\bf B320} (1989) 669.}%
\nref\rey{S. J. Rey and A. Zee, ``Self-Duality Of
Three-Dimensional Chern-Simons Theory,'' Nucl. Phys. {\bf B352}
(1991) 897.}%
 \nref\lr{C. A.
Lutken and G. G. Ross, ``Duality In The Quantum Hall System,''
Phys. Rev. {\bf B45} (1992) 11837,
Phys. Rev. {\bf B48} (1993) 2500.}%
\nref\klz{D.-H. Lee, S. Kivelson, and S.-C. Zhang, Phys. Lett. {\bf 68}
(1992) 2386, Phys. Rev. {\bf B46} (1992) 2223.}%
\nref\luttwo{C. A. Lutken, ``Geometry Of Renormalization Group Flows Constrained By Discrete
Global Symmetries,'' Nucl. Phys. {\bf B396} (1993) 670.}%
\nref\doltwo{B. P. Dolan,  ``Duality And The Modular Group In The
Quantum Hall Effect,'' J. Phys. {\bf A32} (1999) L243,
cond-mat/9805171.}%
\nref\burtwo{C. P. Burgess, R. Dib, and B. P. Dolan, Phys. Rev.
{\bf B62} (2000) 15359,
cond-mat/9911476.}%
 After submitting the original hep-th version of this
paper, I learned that in the context of  fractional quantum Hall
systems, essentially the same definitions were made some time ago,
and a computation similar to that in section 4 was performed, by
Burgess and Dolan \bd. Their motivation came from indications
\refs{\wilczek - \burtwo} of a duality group underlying the
fractional quantum Hall effect.  More generally, Chern-Simons
gauge fields and operations adding and removing them are
extensively used in understanding the quantum Hall effect.  There
is an extensive literature on this; an introduction for particle
physicists is \ref\zee{A. Zee, ``Quantum Hall Fluids,''
cond-mat/9501022.}.

The starting point for the present paper was work by Kapustin and
Strassler \ref\ks{A. Kapustin and M. Strassler, ``On Mirror
Symmetry In Three Dimensional Abelian Gauge Theories,''
hep-th/9902033.} on three-dimensional mirror symmetry \ref\sei{K.
Intriligator and N. Seiberg, ``Mirror Symmetry In
Three-Dimensional Gauge Theories,'' Phys. Lett. {\bf B387} (1996)
512, hep-th/9607207.}.  Kapustin and Strassler considered an
operation that we will call $S$, since it turns out to correspond
to the  generator $\left(\matrix{0 & 1\cr -1 & 0\cr}\right)$ of
$SL(2,{\bf Z})$ that usually goes by that name.  The $S$ operation
is defined as follows. One gauges the $U(1)$ symmetry, introducing
a gauge field $A$ that is coupled to $J$. Thus, if the original
theory has a Lagrangian description in terms of fields $\Phi$ and
a Lagrangian $L$, the new theory has fields $\Phi$ and $A$, and a
Lagrangian that is a gauge-covariant extension of $L$. (In the
simplest case, this extension is just $\tilde L=L+A_iJ^i$.) One
defines a new conformal field theory in which $A$ is treated as a
dynamical field, without adding any kinetic energy for $A$. The
conserved current of the new theory is $\tilde J=*F/2\pi$, where
$F=dA$.

The main result of Kapustin and Strassler was to show that (after
making a supersymmetric extension of the definition) the $S$
operation
  applied to a free hypermultiplet gives back
a mirror free hypermultiplet, and moreover that this implies
mirror symmetry of abelian gauge theories in three dimensions.

$SL(2,{\bf Z})$ is generated along with $S$ by the matrix
$T=\left(\matrix{1 & 1 \cr 0 & 1\cr}\right)$.  The relations they
obey are that $(ST)^3=1$ and that $S^2$ is a central element whose
square is 1 (this is often described by writing $S^2=-1$). In the
present context, $T$ corresponds to a rather trivial operation
that merely shifts the two-point function of the $U(1)$ current
$J$ by a contact term. Though this operation is not very
interesting by itself, it gains interest because it does not
commute with $S$. (This is analogous to four-dimensional gauge
theories, where \ref\rabcard{J. Cardy and E. Rabinovici, ``Phase
Structure Of ${\bf Z}_p$ Models In The Presence Of A Theta
Parameter,'' Nucl. Phys. {\bf B205} (1982) 1; J. Cardy, ``Duality
And The Theta Parameter In Abelian Lattice Models,'' Nucl. Phys.
{\bf B205} (1982) 17.} the simple operation $\theta\to\theta+2\pi$
gains interest because it does not commute with electric-magnetic
duality.) This paper is devoted to understanding from several
points of view that the $S$ and $T$ operations defined for
three-dimensional quantum field theories do generate $SL(2,{\bf
Z})$. \nref\vw{C. Vafa and E. Witten, ``A Strong Coupling Test Of
$S$-Duality,''
Nucl. Phys. {\bf B431} (1994) 3, hep-th/9408074.}%
\nref\witten{E. Witten, ``On $S$ Duality In Abelian Gauge
Theory,'' Selecta Mathematica
{\bf 1} (1995) 383, hep-th/9505186.   }%
\foot{The partition function computed on a three-manifold $Q$
transforms under $SL(2,{\bf Z})$ with a $c$-number phase factor
that depends only on the topology of $Q$ and not on the specific
conformal field theory under discussion, or its currents. This
factor possibly could be removed by modifying the coupling to
gravity or the action of $SL(2,{\bf Z})$ by terms involving the
gravitational background only.  $SL(2,{\bf Z})$ duality of
four-dimensional gauge theory involves a somewhat analogous
topological effect \refs{\vw,\witten}.}

The $T$ operation is closely related to the possibility of having
a Chern-Simons interaction for gauge fields in three dimensions
\ref\djt{S. Deser, R. Jackiw, and S. Templeton, ``Topologically
Massive Gauge Theories,'' Ann. Phys. {\bf 140} (1982) 372.}; in
fact, $T$ simply shifts the Chern-Simons level of the background
gauge field. Kapustin and Strassler in \ks\ considered
Chern-Simons interactions (and also contact terms in two-point
functions) in relation to the $S$ operation.  (In terms of the
quantum Hall effect, $T$ is understood \bd\ as a $2\pi$ shift in
the statistics of the charge carriers.)

\nref\guad{E. Guadagnini, M. Martinelli, and M. Mintchev,
``Scale-Invariant Sigma Models On Homogeneous Spaces,'' Phys.
Lett. {\bf 194B} (1987) 69.}%
\nref\barda{K. Bardacki, E. Rabinovici, and B. Saring, Nucl. Phys.
{\bf B299} (1988) 157.}%
\nref\kar{D. Karabali, Q.-H. Park, H. J. Schnitzer, and Z. Yang,
Phys. Lett. {\bf 216B} (1989) 307; H. J. Schnitzer, Nuc. Phys. {\bf B324} (1989) 412;
D. Karabali and H. J. Schnitzer, Nucl. Phys. {\bf B329} (1990) 649.}%

\nref\app{T. Appelquist and R. D. Pisarski, ``Hot Yang-Mills
Theories And Three-Dimensional QCD,'' Phys. Rev. {\bf D23} (1981)
2305.}%
\nref\jt{R. Jackiw and S. Templeton,  ``How Superrenormalizable
Interactions Cure Their Infrared Divergences,'' Phys. Rev. {\bf
D23} (1981) 2291.}%
\nref\temp{S. Templeton, ``Summation Of Dominant Coupling Constant
Logarithms In QED In Three Dimensions,'' Phys. Lett. {\bf B103}
(1981) 134, ``Summation Of Coupling Constant Logarithms In QED In
Three Dimensions,'' Phys. Rev. {\bf D24} (1981) 3134.}%
\nref\aph{T. Appelquist and U. W. Heinz,``Three-Dimensional $O(N)$
Theories At Large Distances,'' Phys. Rev. {\bf D24} (1981) 2169.}%
\nref\anselmi{D. Anselmi, ``Large $N$ Expansion, Conformal Field Theory, And Renormalization
Group Flows In Three Dimensions,'' JHEP {\bf 0006:042} (2000), hep-th/0005261.}%
 \nref\kp{V. Borokhov, A. Kapustin, and X. Wu,
``Topological Disorder Operators In Three-Dimensional Conformal
Field Theory,''
hep-th/0206054.}%
\nref\newk{V. Borokhov, A. Kapustin, and X. Wu, ``Monopole
Operators And Mirror
Symmetry In Three Dimensions,'' JHEP {\bf 0212} (2002) 044, hep-th/0207074.}%
 Since the definition of $S$ using a gauge field without a kinetic energy seems a bit hazardous
 at first sight, we may derive some encouragement from two dimensions, where
gauging of WZW models without any kinetic energy for the gauge
fields has been used to describe coset models \refs{\guad-\kar}.
 In addition,
there is an illuminating three-dimensional situation in which the
$S$ operation can be made more concrete.  This is the large $N_f$
limit of a theory of $N_f $ free fermions with $U(1)$ symmetry.
The $S$ operation applied to this theory produces a strong
coupling limit of three-dimensional QED that has been much studied
some time ago \refs{\app - \aph} as well as recently
\refs{\anselmi - \newk}.  In this example, the current $J$ has
almost Gaussian correlation functions. In such a situation, we
show that $SL(2,{\bf Z})$ acts by $\tau\to (a\tau+b)/(c\tau+d)$,
where $\tau$ will be defined later in terms of the two-point
function of $J$.  (As noted above, this essentially duplicates
considerations in \bd.)

\nref\brei{P. Breitenlohner and D. Z. Freedman, ``Stability In
Gauged Extended Supergravity,'' Ann. Phys. {\bf 144} (1982)  249.}%
\nref\klebwit{I. R. Klebanov and E. Witten, ``AdS/CFT
Correspondence And Symmetry Breaking,'' Nucl. Phys. {\bf B536}
(1998) 199, hep-th/9905104.}%
We also show that the  $SL(2,{\bf Z})$ symmetry has a closely
related interpretation in the AdS/CFT correspondence. A $U(1)$
global symmetry in three dimensions corresponds to a $U(1)$ gauge
symmetry in a dual description in ${\rm AdS}_4$.  In $U(1)$ gauge
theory in four dimensions, there is an $SL(2,{\bf Z})$ ambiguity
in what we mean by the ``gauge field.''  For each choice, we can
pick an associated boundary condition and, by a familiar
construction, define a conformal field theory on the boundary of
${\rm AdS}_4$ with a conserved current $J$.  The $SL(2,{\bf Z})$
action on conformal field theories on the boundary is induced from
$SL(2,{\bf Z})$ duality transformations on the gauge fields in the
bulk.  This is analogous to the behavior of  scalar fields in AdS
space in a certain range of masses; they can be quantized in two
ways \brei\ leading to two possible CFT duals on the boundary
\klebwit.

In the above summary, we omitted one interesting detail. The
definition of $T$ assumes that we are working on a three-manifold
with a chosen spin structure.  In the absence of a chosen spin
structure, one could define only $T^2$, not $T$, and accordingly
one gets only an action of the subgroup of $SL(2,{\bf Z})$ that is
generated by $S$ and $T^2$.  This is dual to the fact \witten\
that full $SL(2,{\bf Z})$ duality for free abelian gauge fields on
a four-manifold requires a choice of spin structure.

This paper is organized as follows. In section 2, we review some
aspects of abelian Chern-Simons theory in three dimensions (for
more information, see for example \ref\jackiw{R. Jackiw,
``Topological Investigations Of Quantized Gauge Theories,'' in
{\it Current Algebra And Anomalies}, ed. S. B. Treiman et. al.
(World-Scientific, 1985).}), elucidating details such as the role
of a spin structure.  (The rest of the paper can be read while
omitting most details in section 2.) In section 3, we define the
$T$ operation and demonstrate $SL(2,{\bf Z})$ symmetry. In section
4, we consider the case that the current is an almost Gaussian
field. And in section 5, we discuss the duality with gauge fields
in ${\rm AdS}_4$.

\newsec{Abelian Chern-Simons Interactions In Three Dimensions}

\subsec{Generalities}

For an abelian gauge field $A$, let $F=dA$ be the field strength,
and $x=F/2\pi$.  On a compact oriented four-manifold $M$, in
general $\int_M x\wedge x$ is an integer.  If $M$ is spin, then
$\int_M x\wedge x$ is even.

So $J=\int_M F\wedge F/4\pi^2$ is integral in general and is even
if $M$ is spin.\foot{For spin manifolds, a typical example to keep
in mind is $M=T\times T'$ with $T$ and $T'$ being two-tori and
$\int_TF =\int_{T'}F=2\pi$.  For example, if $T$ and $T'$ are made
by identifying boundaries of unit squares in the $x^1-x^2$ and
$x^3-x^4$ planes, respectively, we take $F_{12}=F_{34}=2\pi$ and
other components zero. One readily computes that in this example,
$J=2$. This is the smallest non-zero value of $J$ that is possible
on a spin manifold. For a simple example in which $M$ is not spin
and $J=1$, take $M={\bf CP}^2$ and take $F$ such that
$\int_UF=2\pi$, where $U$ is a copy of ${\bf CP}^1\subset {\bf
CP}^2$.} In physical notation, this would often be written
\eqn\tufo{J={1\over 16\pi^2}\int d^4x\,
\epsilon^{ijkl}F_{ij}F_{kl}.}

Now consider an abelian gauge field $A$ on an oriented
three-manifold $Q$. If $A$ is topologically trivial, the
Chern-Simons functional of $A$ is simply \eqn\bufo{I(A)={1\over
2\pi}\int_Q d^3x\,\epsilon^{jkl}A_j\partial_kA_l.}  It is
important to extend the definition so that it makes sense when $A$
is a connection on a topologically nontrivial line bundle ${\cal
L}$, and hence is not really defined as a one-form.  There is a
standard recipe to do so.  We find a four-manifold $X$ with an
extension of ${\cal L}$ and $A$ over $M$. In four dimensions,
$\epsilon^{ijkl}\partial_i(A_j\partial_kA_l)={1\over
4}\epsilon^{ijkl}F_{ij}F_{kl}$, so we replace \bufo\ by
\eqn\pilop{I_X(A)={1\over 8\pi}\int_X
d^4x\,\epsilon^{ijkl}F_{ij}F_{kl}.} Now we must understand to what
extent this is independent of the choice of $X$.  If $Y$ is some
other four-manifold with an extension of $A$, and $M$ is the
closed four-manifold built by gluing $X$ and $Y$ along their
common boundary $Q$ (after reversing the orientation of $Y$ so the
orientations are compatible), then
\eqn\ilop{I_X(A)-I_Y(A)=2\pi\cdot {1\over
16\pi^2}\int_Md^4x\,\epsilon^{ijkl}F_{ij}F_{kl}=2\pi J(M).}  In
particular, $I_X(A)-I_Y(A)$ is an integer multiple of $2\pi$.

Thus, $\exp(iI(A))$ is independent of the choice of $X$ and the
extension of $A$. This is good enough for constructing a quantum
field theory with $I(A)$ as a term in the action. $I(A)$ is called
the abelian Chern-Simons interaction at level 1.  We often write
it as in \bufo\ even though this formula is strictly valid only
for the topologically trivial case.  (All manipulations we make
later, such as integrations by parts and changes of variables in
path integrals, are easily checked to be valid using the more
complete definition of the Chern-Simons functional.)

If, however, $Q$ is a spin manifold (by which we mean a
three-manifold with a chosen spin structure), we can do better. In
this case, we can pick $X$ and $Y$ so that the chosen spin
structure of $Q$ extends over $X$ and $Y$.  Accordingly, $M$ is
also spin and hence $J(M)$ is an even integer.  Consequently, we
can divide $I(A)$ by two and define \eqn\ufu{\tilde
I(A)={I(A)\over 2}={1\over 4\pi}\int_Q
d^3x\,\epsilon^{jkl}A_j\partial_k A_l,} which is still
well-defined modulo $2\pi$ in this situation. On a
three-dimensional spin manifold, the ``level one-half''
Chern-Simons interaction $\tilde I(A)$ is the fundamental one.

\subsec{ A Trivial Theory}

There are a few more facts that we should know about abelian
Chern-Simons gauge theory in three dimensions.  Consider a theory
with gauge group $U(1)\times U(1)$ and two gauge fields $A$ and
$B$, and with action \eqn\polono{I(A,B)={1\over 2\pi}\int_Q d^3x
\,\epsilon^{jkl}A_j\partial_k B_l.} The extension to a
topologically non-trivial situation in such a way that  $I(A,B)$
is well-defined mod $2\pi$ is made just as we did above for the
case of a single gauge field.  No choice of spin structure is
required here.  This theory has no framing anomaly because the
quadratic form used in writing the kinetic energy has one positive
and one negative eigenvalue; if it is diagonalized, the two fields
with opposite signs of the kinetic energy make opposite
contributions to the anomaly.

We claim, in fact, that this precise theory with two gauge fields
is completely trivial.  One aspect of this triviality is that the
Hamiltonian quantization of the theory is  trivial, in the
following sense: if the theory is quantized on a Riemann surface
$\Sigma $ of any genus, then the physical Hilbert space is
one-dimensional, and the mapping class group of $\Sigma$ acts
trivially on it.  We will call this property Hamiltonian
triviality.

Assuming Hamiltonian triviality for the moment, we will prove
another property that we might call path integral triviality: the
partition function $\hat Z(Q)$ of the theory on an arbitrary
closed three-manifold $Q$ is 1.  (We call this partition function
$\hat Z$ as we will use the name $Z$ for a different partition
fucntion presently.  The partition function is a well-defined
number because the framing anomaly vanishes, as noted above.)
First, for $Q=\S^2\times \S^1$, $\hat Z(Q)$ is the dimension of
the physical Hilbert space on $\S^2$, and so is 1 according to
Hamiltonian triviality.

 Since $\S^2$ can be built by gluing
together two copies of a disc $D$ along their boundary,
$\S^2\times \S^1$ can be built by gluing together two copies $E_1$
and $E_2$ of $D\times \S^1$.  The boundary of $E_1$ is a two-torus
$F=\S^1\times \S^1$. By making a diffeomorphism $S$ of $F$ before
gluing $E_1$ to $E_2$ (the requisite diffeomorphism corresponds to
the modular transformation $S:\tau\to -1/\tau$ exchanging the two
$\S^1$ factors in $F$), one can build $\S^3$.

From these facts we can prove that $\hat Z(\S^3)$ is also 1. In
fact, the path integral on $E_1$ computes a vector $\psi_1$ in the
physical Hilbert space ${\cal H}_F$ of $F$, and the path integral
on $E_2$ likewise computes a vector $\psi_2$ in ${\cal H}_F$.
$\hat Z(\S^2\times \S^1)$ is the overlap $\langle
\psi_1|\psi_2\rangle$, and $\hat Z(\S^3)$ is the matrix element
$\langle\psi_1|\rho(S)\psi_2\rangle$, where $\rho(S)$ is the
linear transformation that represents the diffeomorphism $S$ on
the physical Hilbert space ${\cal H}_F$. The assumption of
Hamiltonian triviality says that $\rho(S)=1$, so $\S^3$ and
$\S^2\times \S^1$ have the same partition function. Hence $\hat
Z(\S^3)=1$.

The same argument can be extended, using standard facts about
three-manifolds, to show that $\hat Z(Q)=1$ for any $Q$. Consider
a genus $g$ Riemann surface embedded in $\S^3$. It has an interior
and also an exterior. They are equivalent topologically and are
called ``handlebodies.''  So $\S^3$ can be made by gluing together
two genus $g$ handlebodies $H_1$ and $H_2$.  Any three-manifold
can be made by gluing together $H_1$ and $H_2$ (for some $g$)
after first making a diffeomorphism $\sigma$ of the boundary of
$H_1$. Assuming Hamiltonian triviality, $\sigma$ acts trivially
and the  argument of the last paragraph shows that the partition
function  of the three-manifold obtained this way is independent
of $\sigma$. Hence all three-manifolds have the same partition
function; in view of the special case $\S^2\times \S^1$, the
common value is clearly 1.

To illustrate, we will compute more explicitly $\hat Z(Q)$ for the
case of a three-manifold $Q$ with $b_1(Q)=0$ (and by Poincar\'e
duality, hence also $b_2(Q)=0$). We simply perform the path
integral \eqn\doint{\int DA\,DB\,\,\exp\left({i\over 2\pi}\int_Q
d^3x \,\epsilon^{jkl}A_j\partial_k B_l\right).} To do the path
integral, we write $A=A_{triv}+A'$, where $A_{triv}$ is a
connection on a trivial line bundle and $A'$ has harmonic
curvature, and similarly $B=B_{triv}+B'$.  In the case at hand, as
$b_2(Q)=0$, $A'$ and $B'$ are both flat and the action is a simple
sum $I(A,B)=I(A_{triv},B_{triv})+I(A',B')$.  The path integral is
a product of an integral over $A_{triv}$ and $B_{triv}$ and a
finite sum over $A'$ and $B'$.

The path integral over $A_{triv}$ and $B_{triv}$ gives a ratio of
determinants in a standard way.  As shown by A. Schwarz in his
early work on topological field theory \ref\aschwarz{A. Schwarz,
``The Partition Function Of A Degenerate Functional,'' Commun.
Math. Phys. {\bf 67} (1979) 1.}, this ratio of determinants gives
$\exp({\cal T})$, where ${\cal T}$ is the Ray-Singer torsion.
Schwarz considered a slightly more general model in which $A$ and
$B$ are twisted by a flat bundle (and one gets the torsion of the
given flat bundle), but in the present instance this is absent, so
we want $\exp({\cal T})$ for the trivial flat bundle.  In three
dimensions, this is equal to $1/\# H_1(Q;{\bf Z})$,\foot{This can
be seen as follows (the argument was supplied by D. Freed). A
three-manifold $Q$ with $b_1=0$ is called a rational homology
sphere and has a cell decomposition with a single 0-cell, $N$
1-cells and 2-cells for some $N$, and a single 3-cell.  The
associated chain complex given by the boundary operator looks like
${\bf Z}\to {\bf Z}^N\to {\bf Z}^N\to {\bf Z}$.  The first and
last differentials vanish (as $b_0=b_3=1$) and after picking an
orientation, $H_0$ and $H_3$ have natural bases (given by a point
in $Q$, and $Q$ itself).  There remains the middle differential;
it is injective, as $b_1=0$, and its cokernel is $H_1(Q;{\bf Z})$.
 The determinant of this differential is thus $|H_1(Q;{\bf Z})|$, and this is
 the exponential of the torsion for the trivial connection, relative to the natural bases
 on $H_0$ and $H_3$.}
 where
for a finite group $F$, $\# F$ denotes the number of elements of
$F$. Poincar\'e duality implies in three dimensions that $\#
H_1(Q;{\bf Z})=\# H^2(Q;{\bf Z})$, a fact that we will use in the
next paragraph.

Now let us perform the sum over $A'$ and $B'$. They are classified
by $x_A,x_B\in H^2(Q;{\bf Z})$, the first Chern classes of the
line bundles on which $A$ and $B$ are connections. However, it is
convenient to break the symmetry between $A$ and $B$ and to look
at $x_B$ differently.  We use the exact sequence $0\to {\bf Z}\to
{\bf R}\to U(1)\to 0$ and the associated exact sequence of
cohomology groups $\dots \to H^1(Q;{\bf R})\to H^1(Q;U(1))\to
H^2(Q;{\bf Z})\to H^2(Q;{\bf R})\to \dots$. Since we are assuming
that $b_1(Q)=b_2(Q)=0$, we have $H^i(Q;{\bf R})=0, $ $i=1,2$ and
hence $H^2(Q;{\bf Z})\cong H^1(Q;U(1))$.  So naturally associated
with $x_B$ is an element $\eta_B\in H^1(Q;U(1))$.  Instead of
summing over $x_A$ and $x_B$ to get the contribution of
non-trivial topologies to the path integral, we can sum over $x_A$
and $\eta_B$.  The sum we want is loosely speaking
\eqn\uccu{\sum_{x_A,\eta_B}\exp\left({i\over 2\pi}\int
d^3x\epsilon^{ijk} A'_i\partial_j B'_k\right),} but here we are
considering topologically non-trivial gauge fields so we need to
use a more precise definition of the action. Actually, the cup
product and Poincar\'e duality give a perfect pairing
$T:H^2(Q;{\bf Z})\times H^1(Q;U(1))\to H^3(Q;U(1))=U(1)$. The
action (for flat gauge fields) can be written in terms of $T$, and
the sum we want is really \eqn\ucon{\sum_{x_A,\eta_B}
T(x_A,\eta_B).} Perfectness of the pairing $T$ says in particular
that for any fixed and nonzero $\eta_B$, we have
$\sum_{x_A}T(x_A,\eta_B)=0$.  On the other hand, if $\eta_B=0$,
then $T=1$ for all $x_A$ and we get $\sum_{x_A}T(x_A,0)=\#
H^2(Q)$. So the sum over topological classes of gauge field gives
a factor $\# H^2(Q)=\# H_1(Q)$, canceling the factor that comes
from the torsion.  Thus, the partition function equals 1, as
claimed.

\bigskip\noindent{\it Another View}

For future reference, we can look at this situation in another
way. The following remarks do not require assuming that
$b_1=b_2=0$ and hold on any three-manifold.   Go back to the path
integral \doint\ and perform first the integral over $A$. The
integral over $A_{triv}$ gives us a delta function setting the
curvature of $B$ to zero. Being flat, $B$ defines an element
$\eta_B\in H^1(Q;U(1))$.  Given that $B$ is flat, the action
depends on $A$ only through its characteristic class $x_A$, and
the sum over $x_A$ (for fixed $B$) is the one encountered in the
last paragraph, $\sum_A T(x_A,\eta_B)$.  As we have just noted,
this sum is a multiple of $\delta(\eta_B)$.  Altogether then, if
we perform first the path integral over $A$, we get a multiple of
$\delta(B)$, that is a delta function saying that $B$ must vanish
up to a gauge transformation.  Since we have found that the
partition function is 1, the multiple of the delta function is
precisely 1. This result is a three-dimensional analog of an a
result used by Rocek and Verlinde \ref\rv{M. Rocek and E.
Verlinde, ``Duality, Quotients, and Currents,'' Nucl. Phys. {\bf
B373} (1992) 630, hep-th/9110053.} in understanding $T$-duality in
two dimensions.

The fact that we have just explained -- the integral over $A$
equals $\delta(B)$ -- is the fact that really will be used in
section 3.  What we gained by the preceeding derivation of
triviality of the theory is the not entirely trivial fact that the
coefficient of $\delta(B)$ is 1.

\bigskip\noindent{\it Hamiltonian Triviality}

To complete the story, it remains to establish Hamiltonian
triviality of the theory. We consider first the case of
quantization on a Riemann surface $\Sigma$ of genus 1. We pick two
one-cycles $C_1$, $C_2$ giving a basis of $H^1(\Sigma;{\bf Z})$.
The gauge invariance and Gauss law constraints imply that the
physical Hilbert space is constructed by quantizing the moduli
space of flat connections mod gauge transformations. (For example,
see \ref\elsei{S. Elitzur, G. Moore, A. Schwimmer, and N. Seiberg,
``Remarks On The Canonical Quantization Of The Chern-Simons-Witten
Theory,'' Nucl. Phys. {\bf B326} (1989) 108.}.) A flat gauge field
$A$ or $B$ is determined mod gauge transformations by
$\alpha_i=\oint_{C_i}A$, $\beta_i=\oint_{C_i}B$. For flat gauge
fields, the action is \eqn\tino{{1\over 2\pi}\int dt
\left(\alpha_1{d\beta_2\over dt}-\alpha_2{d\beta_1 \over
dt}\right).} The symplectic structure is therefore
$\omega=(d\alpha_1 \wedge d\beta_2-d\alpha_2\wedge
d\beta_1)/2\pi$. The $\alpha_i$ and $\beta_j$ all range from $0 $
to $2\pi$, so the phase space volume for each canonically
conjugate pair $\alpha_1,\beta_2$ or $\alpha_2,-\beta_1$ is
$2\pi$.  The quantization of each pair thus leads to precisely one
quantum state.  The physical Hilbert space is obtained by
tensoring together the spaces made by quantizing the commuting
pairs of conjugate variables, and so is also one-dimensional.
Moreover, we can carry out the quantization by regarding the
wavefunctions as functions of (say) the $\beta_i$. Since the
$\beta_i$ are mapped to themselves by modular transformations, the
mapping class group acts in the natural way in this
representation: a modular transformation $\tau$ that maps
$\beta_i$ to $\tau(\beta_i)$ maps a wavefunction $\psi(\beta_i)$
to $ \psi(\tau(\beta_i))$.  (If we carry out the quantization in a
way that is not manifestly modular invariant, the action of the
modular group is more difficult to describe.)

In the $\beta_i$ representation, the unique physical state is a
delta function supported at $\beta_i=0$.  (For example, the
operator $\exp(i\beta_2)$ shifts $\alpha_1$ by $2\pi$ and hence
must act trivially on physical states -- so they have their
support at $\beta_2=0$.  Similarly $\exp(i\beta_1)$ shifts
$\alpha_2$ by $-2\pi$ and must act trivially.) This state is
clearly modular-invariant.  So we have established Hamiltonian
triviality in genus one.

More generally, for $\Sigma$ of genus $g$, we simply have $g$
independent pairs of variables $\alpha_i$, $\beta_j$, each
governed by the same action as above.  The same arguments go
through to show that there is a unique quantum state, given in the
$\beta_j$ representation by a delta function supported at
$\beta_j=0$; the mapping class group acts trivially on this state
for the same reasons as in genus one.  This establishes
Hamiltonian triviality in general.

\subsec{ An Almost Trivial Theory}

Finally, on a spin manifold $Q$, let us consider a $U(1)$ gauge
field $U$ with the level one-half Chern-Simons action
\eqn\levelonehalf{I_U=\tilde I(U)={1\over
4\pi}\int_Qd^3x\,\epsilon^{ijk}U_i\partial_j U_k.} Its partition
function $Z_U$ is a topological invariant of the framed manifold
$Q$ (the framing is needed because of a gravitational anomaly
\ref\witj{E. Witten, ``Quantum Field Theory And The Jones
Polynomial,'' Commun. Math. Phys. {\bf 121} (1989) 351.}) and
changes in phase under a change of framing.  So we cannot expect
to prove that this partition function is 1.  At best we can expect
to show that it is of modulus one.  We will do this first from a
Hamiltonian point of view, and then from a path integral point of
view.

From the Hamiltonian point of view, the claim is that on a Riemann
surface $\Sigma$ of any genus, the physical Hilbert space of this
theory is one-dimensional.  However, there is no claim that the
mapping class group acts trivially.  Indeed, because of the
framing anomaly, it is really a central extension of the mapping
class group that acts naturally; we will not describe this action
here.

To see that the physical Hilbert space is one-dimensional, we
simply proceed as above.  On a surface $\Sigma$ of genus 1, the
action, evaluated as before for flat connections modulo gauge
transformations, comes out to be \eqn\naughty{{1\over 2\pi}\int
dt\, \alpha_1{d\alpha_2\over dt}.} Again, the phase space volume
is $2\pi$, and there is one quantum state.  But there is no
manifestly modular-invariant way to carry out the quantization,
and hence the above argument for triviality of the action of the
mapping class group does not apply.  One can take the
wavefunctions to be functions of $\alpha_1$, or functions of
$\alpha_2$, or of a possibly complex linear combination thereof,
but no such choice is invariant under the action of the mapping
class group, so no such choice enables one to simply read off how
the mapping class group acts.

Since the theory is unitary, the mapping class group acts only by
phases -- complex numbers of modulus one.  When we try to compute
$Z_U(Q)$ for general $Q$ by cutting and pasting, using the
arguments we used to compute $\hat Z(Q)$, everything is as before
except that we run into undetermined phases in the action of the
mapping class group. (These phases could be analyzed, but we will
not do so.)  So we cannot argue that $Z_U(Q)$ equals 1. We can
only argue that it is of modulus one.

Now we will see how to reach the same conclusion from a path
integral point of view. Let $V$ be another $U(1)$ gauge field with
a Chern-Simons action of level minus one-half,
\eqn\minlevl{I_V=-\tilde I(V)=-{1\over
4\pi}\int_Qd^3x\,\epsilon^{ijk}V_i\partial_jV_k.} The partition
functions $Z_U=\int DU\,\exp(i\tilde I(U))$ and $Z_V=\int
DV\,\exp(-i\tilde I(V))$ are complex conjugates of one another,
$Z_V=\overline Z_U$, since (after replacing $U$ by $V$) the
integrands of the path integrals are complex conjugates.

On the other hand, we claim that $Z_UZ_V=1$.  The product $Z_UZ_V$
is the partition function of the combined theory with action
\eqn\combac{I(U,V)={1\over 4\pi}\int_Q
d^3x\epsilon^{ijk}(U_i\partial_jU_k-V_i\partial_jV_k).} Now make
the change of variables $V\to B=U+V$ with $U$ fixed.  The action
becomes \eqn\tombac{I(B,V)={1\over 2\pi}\int_Q
d^3x\epsilon^{ijk}U_i\partial_jB_k-{1\over
4\pi}\int_Qd^3x\epsilon^{ijk}B_i\partial_jB_k.} Performing first
the path integral over $U$ gives (as we saw in section 2.2) a
delta function setting $B=0$.  This means that the path integral
is unaffected if we drop the $BdB$ term.  Hence $Z_UZ_V$ is equal
to the partition function $\hat Z=1$ that was found in section 2.2
for the theory in which the $BdB$ term is omitted from the action.
So $Z_UZ_V=1$, as we aimed to prove, and hence $|Z_U|^2=1$.

It is conceivable that there is some natural way to pick a framing
that would make $Z_U=1$.  In that case, the level one-half theory,
understood with this framing, would be trivial.

\newsec{Action Of $SL(2,{\bf Z})$ On Conformal Field Theories}

In this section, we first describe the operation $S$ that was used
in \ks\ to describe three-dimensional mirror symmetry. Then we
describe an additional operation, which we will call $T$, and show
that $S$ and $T$ together generate $SL(2,{\bf Z})$.

The objects we will study will be conformal field theories in
three spacetime dimensions with a global $U(1)$ symmetry. The
$U(1)$ symmetry is generated by a conserved current $J$.  However,
we need to be more precise in several ways.

First of all, we regard the choice of $J$ as part of the
definition of the theory.  The current $-J$ would also generate
the $U(1)$ symmetry.  It turns out that the central element $-1\in
SL(2,{\bf Z})$ is represented by the operation $J\to -J$, leaving
the theory otherwise fixed.

Second, to be more precise, what we study will be a conformal
field theory with a choice of $J$ and a precise definition of the
$n$-point functions of $J$.  The reason that we make this last
request is that in three dimensions, it is possible to have a
conformally invariant contact term in the two-point function of a
conserved current, \eqn\moglo{\langle J_k(x)J_l(y)\rangle \sim
{w\over 2\pi}\epsilon_{jkl} {\partial\over \partial
x^j}\delta^3(x-y)+\dots} There is in general no natural way to fix
the coefficient $w$. (Shifts in $w$ were encountered in \ks\ in
some examples.)  We regard specification of $w$ as part of the
definition of the theory. $T$ will act essentially by shifting
$w$.

This is still an imperfect description of the type of object we
want to study.  To be a little more precise, we introduce an
auxiliary gauge field $A$ and consider the generating functional
of correlation functions of $J$, which we provisionally take to be
$\left\langle \exp\left(i\int_Qd^3x\,
A_iJ^i\right)\right\rangle.$\foot{We actually consider
unnormalized correlation functions; that is, we do not divide by
the value at $A=0$.} It is convenient but not necessary to assume
that our theory has a Lagrangian description with fields $\Phi$
and a Lagrangian $L(\Phi)$.  In that case, the generating
functional can be provisionally represented \eqn\huto{\left\langle
\exp\left(i\int_Q d^3x A_iJ^i\right)\right\rangle=\int
D\Phi\,\exp\left(i\int_Qd^3x(L(\Phi)+A_iJ^i)\right),} where the
path integral is carried out only over $\Phi$, with $A$ being a
spectator, a background gauge field.

However, we wish to modify the definition of the generating
functional so that it is invariant under gauge transformations of
$A$.  This will often but not always be the case with the
definition we have given so far. A familiar counterexample arises
if $\Phi$ is a complex scalar field, $J$ the current that
generates the $U(1)$ symmetry $\Phi\to \exp(i\theta)\Phi$, and
$L(\Phi)=|\partial\Phi/\partial x^i|^2$. In this example, the
current $J$, though conserved, is not invariant under local gauge
transformations.  The gauge-invariant generalization of $L(\Phi)$
is not $L(\Phi)+A_iJ^i$, but $\tilde L(\Phi,A)=|D\Phi/Dx^i|^2$,
where $D/Dx^i=\partial/\partial x^i+iA_i$.  This includes a term
quadratic in $A$.  (From a general conformal field theory point of
view, this extra term is needed because of an additional primary
operator -- in this case $|\Phi|^2$ -- that appears in the
operator product expansion of two currents.)

So finally we come to the precise definition of the class of
objects that we really want to study.  What we really want is a
three-dimensional conformal field theory with a choice of
gauge-invariant quantum coupling to a background $U(1)$ gauge
field $A$. In case the conformal field theory has a Lagrangian
description, this means that we are given a gauge-invariant and
conformally invariant extension $\tilde L(\Phi,A)$ of the original
Lagrangian, and we can define the gauge-invariant functional
\eqn\tommy{\exp(i\Gamma(A))=\int D\Phi \exp\left(i\int
d^3x\,\tilde L(\Phi,A)\right).} The coupling to the background
gauge field is required to be gauge-invariant at the quantum
level. Picking such a gauge-invariant coupling entails in
particular, as we explain later, a choice of the Chern-Simons
coupling for the background gauge field.\foot{The need to make
such a choice is particularly clear in case there is a parity
anomaly \ref\redlich{N. Redlich, ``Parity Violation And Gauge
Non-Invariance Of The Effective Gauge Field Action In Three
Dimensions,'' Phys. Rev. {\bf D29} (1984) 2366.} in the coupling
of the conformal field theory to the background gauge field; in
this case, the Chern-Simons coupling cannot be zero as it is in
fact a half-integral multiple of the level one-half functional
$\tilde I(A)$.}

\bigskip\noindent{\it The $S$ Operation}

Now we can define the $S$ operation. Roughly speaking, instead of
regarding $A$ as a background field, we now regard $A$ as a
dynamical field, and perform the path integral over $A$ as well as
$\Phi$.  We thus define a ``dual'' theory whose fields are $A$ and
$\Phi$ and whose Lagrangian is $\tilde L(\Phi,A)$.

However, for the dual theory to be of the same type that we have
been considering, we must define a conserved current in this
theory and explain how to couple it to a background gauge field.
We define the conserved current of the dual theory to be $\tilde
J_i=\epsilon_{ijk}F_{jk}/4\pi=\epsilon_{ijk}\partial_jA_k/2\pi$;
it is conserved because of the Bianchi identity obeyed by $F$. We
denote the background gauge field of the dual theory as $B$. The
current $\tilde J_i$ is gauge-invariant as well as conserved, so a
gauge-invariant coupling to the background field $B$ is made
simply by adding a new interaction $\tilde J^i B_i$. The combined
Lagrangian is therefore simply \eqn\ungo{L'(\Phi,A,B)={1\over
2\pi}\epsilon^{ijk}B_i\partial_jA_k +\tilde L(\Phi,A).} This
theory, with $\Phi$ and $A$ understood as dynamical fields and $B$
as a background gauge field, is the one we obtain by applying the
$S$ operation to the original theory.  This is the definition of
$S$.\foot{Defining the path integral for a gauge field in three
dimensions can in general require a choice of framing of the
three-manifold \witj.  It is at this step that there appears the
potential for a $c$-number gravitational effect in the action of
$SL(2,\Z)$.}

Now, following \ks, we want to compute $S^2$. We apply $S$ a
second time by making the background gauge field $B$ dynamical and
adding a new spectator gauge field $C$, coupled this time to the
current $\tilde J_i(B)=\epsilon_{ijk}\partial_jB_k/2\pi$ made from
$B$. So the theory obtained by applying $S$ twice has dynamical
fields $\Phi$, $A$, and $B$, background gauge field $C$, and
Lagrangian \eqn\tungo{L''(\Phi,A,B,C)={1\over
2\pi}\epsilon^{ijk}C_i\partial_jB_k+ {1\over
2\pi}\epsilon^{ijk}B_i\partial_jA_k +\tilde L(\Phi,A).}

After an integration by parts, the part of the action that depends
on $B$ is \eqn\yungo{{1\over 2\pi}\int d^3x\,
\epsilon_{ijk}B_i\partial_j(A+C)_k.} The integral over $B$ is
therefore very simple.  As explained near the end of section 2.2,
it simply gives a delta function $\delta(A+C)$ setting $A+C$ to
zero, up to a gauge transformation.  The integral over $A$ is
therefore also trivial; it is carried out by setting $A=-C$. After
integrating out $A$ and $B$, we therefore get a theory with
dynamical field $\Phi$, background gauge field $C$, and Lagrangian
$\tilde L(\Phi,-C)$. This is just the original theory with the
sign of the current reversed.  So this justifies the assertion
that the effect of applying $S^2$ is to give back the original
theory with the sign of the current reversed.  We write this
relation as $S^2=-1$, where $-1$ leaves the theory unchanged and
reverses the sign of the current.

\bigskip\noindent{\it The $T$ Operation}

Now we want to define another operation that we will interpret as
the second generator $T$ of $SL(2,{\bf Z})$.

The operation will act on conformally invariant theories with
dynamical fields $\Phi$, background fields $A$, and Lagrangian
$\tilde L(\Phi,A)$.  We simply exploit the lack of uniqueness in
passing from the underlying conformally invariant Lagrangian
$L(\Phi)$ to its gauge-invariant extension $\tilde L(\Phi,A)$.

What lack of uniqueness is there?  For the present purposes, we
want to change $\tilde L(\Phi,A)$ only by terms that vanish at
$A=0$; other terms represent moduli of the conformal field theory
that we started with, rather than ambiguities in the coupling to a
background gauge field.

There are in fact no locally gauge-invariant operators vanishing
at $A=0$ that can be added to $\tilde L(\Phi,A)$ while preserving
conformal invariance.  For example, a Lorentz-invariant functional
of $A$ only would have at least dimension four, the lowest
dimension possibility being the usual gauge action $F_{ij}F^{ij}$.
A locally gauge-invariant coupling of a gauge field to the $\Phi$
field that vanishes at $F=0$ must involve at least one explicit
factor of $F$; the case of lowest dimension is an interaction
$\epsilon^{ijk}F_{ij}X_k$ with $X_k$ some conformal field made
from $\Phi$.  Unitarity implies that in three dimensions a
vector-valued conformal field such as $X$ has dimension greater
than 1, so this interaction again spoils conformal invariance.

The only remaining option is to add to $\tilde L(\Phi,A)$ the
Chern-Simons interaction.  We add the Chern-Simons interaction at
level one-half: \eqn\toggof{\tilde L(\Phi,A)\to \tilde
L(\Phi,A)+{1\over 4\pi}\epsilon^{ijk}A_i\partial_j A_k.} The term
we have added is not locally gauge-invariant, but it is
gauge-invariant up to a total derivative; more to the point, as
reviewed in section 2, its integral over a three-manifold $Q$ with
a chosen spin structure is gauge-invariant and well-defined mod
$2\pi$.

What we will call the $T$  operation consists of adding to the
Lagrangian the Chern-Simons coupling of the background gauge field
$A$. This operation is essentially trivial, in that the term which
is added depends only on the background field  and not on the
dynamical field $\Phi$. The effect of the $T$ operation on the
generating functional of current correlation functions (or its
generalization \tommy) is \eqn\jommo{ \left\langle
\exp\left(i\int_Q d^3x\,A_iJ^i\right)\right\rangle\to
 \left\langle \exp\left(i\int_Qd^3x\,
A_iJ^i\right)\right\rangle\exp\left({i\over 4\pi}\int
d^3x\,\epsilon^{ijk}A_i\partial_j A_k\right).} This is equivalent
to adding to the two-point function of $J$ a contact term of the
form described in \moglo, with a definite coefficient. Thus, the
theory transformed by $T$ is the same as the original theory  but
with a contact term added to the correlation functions.

As we have reviewed in section 2, if we do not want to endow $Q$
with a spin structure, we must double the Chern-Simons coupling in
\toggof.  This means that without using a spin structure, we can
only define the operation $T^2$ and not $T$.

For our purposes in the present paper, the reason that the trivial
operation $T$ is worth discussing is that it does not commute with
$S$.  Let us, for practice, work out $ST$ and compare it to $TS$.

To compute $ST$, we first act with $T$ by coupling to a background
gauge field $A$ and adding the Chern-Simons coupling of $A$ at
``level one-half.''  Then we act with $S$ by making $A$ dynamical,
and adding a background gauge field $B$ that has a coupling to the
current $\tilde J_i(A)=\epsilon_{ijk}\partial_jA_k/2\pi$.  By the
time all this is done, we have the Lagrangian
\eqn\pijno{L_{ST}(\Phi,A,B) ={1\over
2\pi}\epsilon_{ijk}B_i\partial_jA_k +{1\over
4\pi}\epsilon_{ijk}A_i\partial_jA_k+\tilde L(\Phi,A),} with the
dynamical fields being $\Phi$ and $A$.

To instead   compute $TS$, we  first act with $S$ by making the
background field $A$ dynamical and  including a background gauge
field $B$ that couples to $\tilde J$.  Then we  act with $T$ by
adding the level one-half Chern-Simons coupling of $B$.  We get
the Lagrangian \eqn\ginjo{L_{TS}(\Phi,A,B) ={1\over
4\pi}\epsilon_{ijk}B_i\partial_jB_k+{1\over
2\pi}\epsilon_{ijk}B_i\partial_jA_k +\tilde L(\Phi,A).} The
theories with Lagrangian $L_{TS}$ and with $L_{ST}$ are not
equivalent, since the Lagrangians are different and cannot be
transformed into one another by a change of variables.  (Though
this fact does not affect the answer, note that, as $B$ is a
background field, we should only consider changes of variable that
leave $B$ fixed.)

\bigskip\noindent{\it $SL(2,{\bf Z})$ Action}

Now that we have some practice with such computations, let us try
to prove that $(ST)^3=1$.  (Actually, we will see that $(ST)^3=1$
modulo a $c$-number gravitational correction, a topological
invariant that depends only on the manifold $Q$ and not on the
specific theory under discussion.) Together with the
 result $S^2=-1$, this will show that $S$ and
$T$ together generate $SL(2,{\bf Z})$.

To act with $ST$, we add a level one-half Chern-Simons coupling
for the background gauge field $A$, make that field dynamical, and
add a new background gauge field $B$, coupled to $\tilde J_i(A)=
\epsilon_{ijk}\partial_jA_k/2\pi$.  To act with $ST $ again, we
add a level one-half Chern-Simons interaction of $B$, make $B$
dynamical, and add a new background gauge field $C$ coupled to
$\tilde J_i(B)$.  Finally, to act with $ST$ a third time, we add a
level one-half Chern-Simons interaction for $C$, make $C$
dynamical, and add a new background gauge field $D$ coupled to
$\tilde J_i(C)$. All told, the Lagrangian after acting with
$(ST)^3$ is \eqn\rygo{\eqalign{L_{(ST)^3}(\Phi,A,B,C,D)= &{1\over
2\pi}\epsilon_{ijk}D_i\partial_jC_k +{1\over
4\pi}\epsilon_{ijk}C_i\partial_jC_k+{1\over
2\pi}\epsilon_{ijk}C_i\partial_jB_k \cr &+{1\over
4\pi}\epsilon_{ijk}B_i\partial_jB_k+{1\over
2\pi}\epsilon_{ijk}B_i\partial_jA_k +{1\over
4\pi}\epsilon_{ijk}A_i\partial_jA_k+\tilde L(\Phi,A).\cr}} The
dynamical fields are $\Phi,A,B,$ and $C$; $D$ is a background
field.

To analyze this theory, we simply replace $C$ by a new variable
$\tilde C=B+C+D$, leaving $\Phi$, $A,$ $B$, and $D$ fixed.  The
Lagrangian becomes \eqn\tygo{\eqalign{L_{(ST)^3}(\Phi,A,B,\tilde
C,D)= &{1\over 2\pi}\epsilon^{ijk}B_i\partial_j(A_k-D_k)+{1\over
4\pi}\epsilon^{ijk}\tilde C_i\partial_j\tilde C_k \cr &+{1\over
4\pi}\epsilon^{ijk}A_i\partial_j A_k-{1\over
4\pi}\epsilon^{ijk}D_i\partial_jD_k+\tilde L(\Phi,A).\cr}} We
perform first the path integral over $B$, which, as explained in
section 2.2, gives a delta function setting $A-D=0$ up to a gauge
transformation.  We next perform the path integral over $A$ simply
by setting $A=D$. We reduce to \eqn\ygo{L_{(ST)^3}(\Phi,\tilde
C,D) =\tilde L(\Phi,D) +{1\over 4\pi}\epsilon^{ijk}\tilde
C_i\partial_j\tilde C_k.} Thus, apart from the decoupled $\tilde
C$ theory, the operation of $(ST)^3$ gives us back the original
theory coupled to a background gauge field $D$ with the original
Lagrangian $\tilde L(\Phi,D)$. In this sense $(ST)^3=1$.

The $\tilde C$ theory is a Chern-Simons theory at level one-half
and was analyzed in section 2.3. It multiplies the partition
function by a complex number of modulus one that is a topological
invariant, independent of the specific conformal field theory
under study and decoupled from both the theory and its currents.
Our analysis in this paper is really not precise enough to give
the best way of dealing with topologically invariant $c$-number
contributions that depend on the gravitational background only.
(It may be that by redefining $T$ by a topological invariant, one
can avoid the gravitational correction to $(ST)^3=1$.  This can be
done if the Chern-Simons theory at level one-half is isomorphic to
the cube of some other topological field theory; one would then
modify the definition of $T$ to include tensoring with the dual of
this theory.  It may even be that the level one-half Chern-Simons
theory, or its cube, is trivial with a natural choice of framing.)

\newsec{Modular Action On Current Two-Point Function}

The $S$ operation can be made much more explicit in the case
\refs{\app-\newk} of $N_f$ free fermions with $U(1)$ symmetry for
large $N_f$.  After coupling to a background gauge field $A$, we
have \eqn\ingo{L=\int d^3x\,\sum_{i=1}^{N_f}\bar\psi_ii\Gamma\cdot
D\psi_i.} Upon integrating out the $\psi_i$, we get an effective
action for the gauge field that takes the form $\int d^3x
\,N_f\left(F_{ij}\Delta^{-1/2}F^{ij}+\dots\right)$, where $\Delta
$ is the Laplacian, and the ellipses refer to terms of higher
order in the gauge field strength $F$.  This theory can be
systematically studied for large $N_f$  as it is weakly coupled,
with the effective cubic coupling (after absorbing a factor of
$1/\sqrt{N_f}$ in $F$ so that the kinetic energy is of order one)
being proportional to $1/\sqrt{N_f}$.

The large $N_f$ theory has the property that, before or after
acting with $S$, the current has nearly Gaussian correlation
functions. Other examples in which the current is nearly Gaussian
come from the AdS/CFT correspondence, which we consider in section
5.

In this section, we will analyze the action of  $SL(2,{\bf Z})$ on
the two-point function of the $U(1)$ current, for the case that
the current is nearly Gaussian so that its correlations are
characterized by giving the two-point function.  We will see that
from the current two-point function we can define a complex
parameter $\tau$, valued in the upper half plane, on which
$SL(2,{\bf Z})$ acts in the familiar fashion $\tau\to
(a\tau+b)/(c\tau+d)$, with $a,b,c$, and $d$ integers, $ad-bc=1$.

The general form of the current two-point function in momentum
space in a three-dimensional conformal field theory is
\eqn\notang{ \left\langle J_i(k)J_j(-k)\right\rangle
=\left(\delta_{ij}k^2-k_ik_j\right){t\over 2\pi\sqrt{k^2}}
+\epsilon_{ijr}k_r{w\over 2\pi}.} Here $t$ and $w$ are constants
and $t$ is positive if $J_i$ is hermitian.
 To carry out the
transform by $S$, we couple the current to a gauge field $A_i$. In
momentum space, the effective action for $A_i$, after including
gauge-fixing terms that cancel $k_iA^i $ terms, is
\eqn\rotang{\int d^3k\left({t\over
4\pi}A_i(k)\sqrt{k^2}A_i(-k)+{w\over
4\pi}\epsilon_{ijr}A_i(k)k_rA_j(-k)\right).} The propagator of
$A_i$ is $\langle A_i(k)A_j(-k)\rangle=N_{ij}$, where $N_{ij}$ is
the inverse of  the matrix \eqn\yurgo{M_{ij}={t\over
2\pi}\sqrt{k^2}\delta_{ij}+{w\over 2\pi}\epsilon_{ijr}k_r.} The
inverse is \eqn\purgo{N_{ij}={2\pi\delta_{ij}\over
\sqrt{k^2}}{t\over t^2+w^2}-{2\pi\epsilon_{ijr}k_r\over
k^2}{w\over t^2+w^2}+{2\pi k_ik_j\over (k^2)^{3/2}}{w^2\over t(
t^2+w^2)}.} The current of the theory transformed by $S$ is
$\tilde J_i=\epsilon_{ijr}\partial_j A_r/2\pi$, or in momentum
space $\tilde J_i(k)=-(i/2\pi)\epsilon_{ijr}k_jA_r(k)$.  Using
this and the propagator of $A_i$, one determines that the
two-point function of $\tilde J$ is \eqn\torango{\langle\tilde
J_i(k)\tilde J_j(-k)\rangle = {1\over
2\pi\sqrt{k^2}}\left(\delta_{ij}k^2-k_ik_j\right){t\over t^2+w^2}
-{\epsilon_{ijr}k_r\over 2\pi}{w\over t^2+w^2}.} Comparing to
\notang, we see that $\tau=w+it$, which takes values in the upper
half plane, has transformed via $\tau\to -1/\tau$.

The $T$ transformation of three-dimensional conformal field
theories was defined to shift the two point function of $J$ by
$w\to w+1$.  This amounts to $\tau\to \tau+1$. The results for the
transformation of $\tau$ under $S$ and $T$ can be summarized by
saying that $\tau$ transforms under $SL(2,{\bf Z})$ in the
standard way $\tau\to (a\tau+b)/(c\tau+d)$.

\newsec{Interpretation In AdS/CFT Correspondence}

Finally, in this section, we discuss this $SL(2,{\bf Z})$ action
on three-dimensional conformal field theories in the light of the
AdS/CFT correspondence.

Consider a four-dimensional gravitational theory with negative
cosmological constant and an unbroken $U(1)$ gauge group.  For
example, if the true cosmological constant in the physical vacuum
is negative, then the real world is described by such a theory,
with the $U(1)$ being that of electromagnetism.

Now let us consider constructing a dual three-dimensional
conformal field theory.  We let $A$ denote the massless gauge
field in the four-dimensional bulk.  We denote four-dimensional
Anti de Sitter space, with (Euclidean signature) metric
\eqn\rump{ds^2={L^2\over z^2}\left(dz^2+d\vec x^2\right),} as $X$.
Its conformal boundary $Y$ is at $z=0$ and has coordinates $\vec
x$.

The standard construction is to fix a gauge field $\vec A$ on $Y$
and perform the path integral on $X$ with a boundary condition
requiring  that in the limit $z\to 0$, the part of $A$ tangent to
the boundary is equal to $\vec A$.   The path integral with these
boundary conditions is then interpreted as computing the
generating functional $\left\langle\exp(i\int_Yd^3x\,\vec A\cdot
\vec J)\right\rangle$ of current correlators in the boundary
conformal field theory; here $\vec J$ is the conserved current of
the boundary theory. (In this section only, we write $\vec A$,
$\vec J$ for three-dimensional gauge fields and currents.)

In particular, if one does not want to insert any currents at all
on the boundary, one asks simply that, for $z\to 0$, the
tangential part of $A$ should vanish up to a gauge transformation.
This can be described in a gauge-invariant language by saying that
the magnetic field $\vec B$ (defined as usual by $\vec
B_i=\epsilon_{ijk}\partial_j\vec A_k$) vanishes on $Y$. More
generally, with currents inserted, the recipe is to compute the
bulk partition function as a function of a specified choice of
$\vec B$ (or, equivalently but more conveniently, a specified
$\vec A$).

This recipe is clearly not invariant under $SL(2,{\bf Z})$ duality
of the four-dimensional $U(1)$ theory. For example, the $S$
transformation in $SL(2,{\bf Z})$ maps $\vec B\to \vec E$, $\vec
E\to -\vec B$, and maps the low energy gauge field $A$ to a dual
gauge field $A'$. Applying the standard AdS/CFT recipe in terms of
$A'$ is equivalent in terms of the original gauge field $A$ to
using a boundary condition $\vec E=0$ instead of $\vec B=0$ (and
more generally, computing current correlators by varying the
boundary values of $\vec E$ instead of the boundary values of
$\vec B$).

More generally, we could make an arbitrary $SL(2,{\bf Z})$
transformation to introduce a new gauge field before applying the
standard AdS/CFT recipe.  In terms of the original gauge field,
this corresponds to a boundary condition setting to zero a linear
combination of $\vec E$ and $\vec B$.  So altogether, depending on
the choice of boundary conditions, an AdS theory with a $U(1)$
gauge field in four dimensions has infinitely many possible CFT
duals on the boundary.

To get another view of the problem, let us consider the  AdS/CFT
correspondence in a spacetime with Lorentz signature. To do so, we
replace $d\vec x^2$ in \rump\ by $dx^2+dy^2-dt^2$, with $t,x,y$
being coordinates of three-dimensional Minkowski spacetime.  In
fact, we want to work with the completed ${\rm AdS}_4$ spacetime,
whose boundary has topology $\S^2\times \R$ ($\S^2$ parametrizes
space and $\R$ parametrizes time); the Lorentzian continuation of
\rump\ describes only a ``Poincar\'e patch'' of this.

The four-dimensional bulk theory has both electric and magnetic
charges.  (In the case of the real world, we cannot claim that the
magnetic charges are an experimental fact!)  This might lead one
to expect that the boundary theory should have $U(1)\times U(1)$
global symmetry, but actually in the standard AdS/CFT
correspondence, a massless gauge field in bulk leads to one $U(1)$
symmetry on the boundary, not two.

It is not hard to see what happens.  In one construction, we
require (in the absence of operator insertions on the boundary)
that  $\vec B=0$ on the boundary.\foot{Here, as explained above,
$\vec B$ is the part of the four-dimensional curvature that is
tangent to the boundary and so involves $dx_i\wedge dx_j$ where
$x_i=x,y,$ and $t$.  The other components $\vec E$ are $dx_i\wedge
dz$ where $z$ is the normal coordinate. This decomposition is
natural in the AdS/CFT correspondence, but is non-standard in the
sense that usually such a decomposition is made using the time
coordinate $t$ rather than the normal coordinate $z$.} In this
case, magnetic charge is forbidden, since a state with a net
magnetic charge in the bulk would have an inescapable $\vec B$ on
the boundary. So the conserved quantity of the boundary theory
corresponds to electric charge in bulk.

Alternatively, suppose that the boundary condition, in the absence
of operator insertions on the boundary, is $\vec E=0$.  Now, net
electric charge in bulk is forbidden, but there is no problem with
having a net magnetic charge.  The net magnetic charge corresponds
to the conserved quantity in the boundary theory.

More generally, after making an arbitrary $SL(2,{\bf Z})$
transform of the boundary condition, only one linear combination
of net electric and magnetic charge is allowed in the bulk, and
corresponds to the conserved charge of the boundary theory.

There is no claim here, just as there was none in the earlier
sections of this paper, that the different theories obtained with
different boundary conditions on the gauge field are equivalent.
This may be so in some special cases, but in general the
$SL(2,{\bf Z})$ duality symmetry of the low energy theory
transforms one boundary condition, and one boundary conformal
field theory, to another inequivalent one.

In the remainder of this section, we analyze the $S$ and $T$
duality operations of the bulk theory, and argue that they
correspond on the boundary to the operations of the same names
that we defined in section 3 for three-dimensional conformal field
theories with $U(1)$ symmetry.

\bigskip\noindent{\it The $T$ Operation}

In abelian gauge theory in four dimensions, the generator $T$ of
$SL(2,{\bf Z})$ corresponds to a $2\pi$ shift in the theta angle.
Let us see what this operation corresponds to in the boundary
conformal field theory.

The $\theta$-dependent term in the action of four-dimensional
abelian gauge theory is \eqn\huggot{I_\theta={\theta\over
32\pi^2}\int d^4x\, \epsilon^{ijkl}F_{ij}F_{kl}.} On a closed
four-manifold $X$, the change in $I_\theta$ under
$\theta\to\theta+2\pi$ is $\pi J$, where $J$ (defined in section
2.1) is even on a spin manifold.  So on a closed spin manifold,
$\exp(iI_\theta)$ is invariant under $\theta\to\theta+2\pi$.  (On
a closed four-manifold that is not spin, the symmetry is
$\theta\to\theta+4\pi$.  For background on duality symmetry of
abelian gauge theory on a four-manifold, showing that complete
$SL(2,{\bf Z})$ symmetry holds only on a spin manifold, see
\witten. Of course, a theory -- like the one describing the real
world -- that contains neutral fermions like neutrinos can only be
formulated on spin manifolds.)

On a four-manifold $X$ with boundary $Y$, the change in $I_\theta$
under $\theta\to\theta+2\pi$ is still given by the same expression
$\pi J=(1/16\pi)\int d^4x\, \epsilon^{ijkl}F_{ij}F_{kl}$. But this
is no longer an integral multiple of $2\pi$.  Rather, as we
reviewed in section 2, it differs from being an integral multiple
of $2\pi$ by a functional of the boundary values of the gauge
field which is precisely the level one-half Chern-Simons
functional $\tilde I(\vec A)$. For topologically trivial gauge
fields, this functional can be written \eqn\pugggot{\tilde I(\vec
A)={1\over 4\pi}\int d^3x \,\epsilon^{ijk}\vec A_i
\partial_j\vec A_k.}
(We follow the convention in this section of letting $\vec A$
denote the restriction of the connection $A$ to the boundary.)

Thus, under $\theta\to\theta+2\pi$, the theta dependent factor
$\exp(iI_\theta)$ of the integrand of the path integral is
multiplied by $\exp(i\tilde I(\vec A))$.  Hence, under
$\theta\to\theta+2\pi$, the path integral $Z_{\vec A}$ computed
with specified boundary values $\vec A$ for the gauge field
transforms as \eqn\transfo{Z_{\vec A}\to Z_{\vec A}\exp(i\tilde
I(\vec A)).}

In the boundary conformal field theory, $Z_{\vec A}$ is
interpreted as  the generating function $\left\langle\exp(i\int
d^3x\, \vec A\cdot \vec J)\right\rangle$ of current correlation
functions (or its generalization \tommy). So in view of \transfo,
this generating function transforms under $\theta\to\theta+2\pi$
as \eqn\ransfo{\left\langle\exp\left(i\int d^3x\, \vec A\cdot \vec
J\right)\right\rangle\to \left\langle\exp\left(i\int d^3x\, \vec
A\cdot \vec J\right)\right\rangle \exp\left(i\tilde I(\vec
A)\right).} But this transformation law for the current
correlation functions is the definition of the $T$ operation in
the boundary conformal field theory.  So the
$\theta\to\theta+2\pi$ operation in the bulk theory does induce
the operation that we have called $T$ in the boundary theory.

\bigskip\noindent{\it An Analogy For The $S$ Generator}

 The generator $S$ of $SL(2,\Z)$ exchanges electric and
magnetic fields, so it  corresponds from the AdS point of view to
replacing the boundary condition $\vec B=0$ (or a generalization
of this in which $\vec B$ is specified  to compute current
correlators) with $\vec E=0$ (or a generalization in which $\vec E
$ is specified). We want to show that this induces on the boundary
the $S$ operation. The discussion will not have the degree of
precision that we attained above for $T$.

First, we will treat an analogous problem. We consider a scalar
field in $(d+1)$-dimensional Anti de Sitter space with mass $m^2$.
The Euclidean action is \eqn\toffo{L={1\over
2}\left((\nabla\phi)^2+m^2\phi^2\right).} The general solution
behaves near $z=0$ -- that is, near the boundary of Anti de Sitter
space -- as \eqn\noffo{\phi(z,\vec x)=z^{\Delta_+}\alpha(\vec
x)+z^{\Delta_-}\beta(\vec x),} where $\Delta_+$ and
$\Delta_-<\Delta_+$ are the two roots of the quadratic equation
$\Delta(\Delta-d)=m^2L^2$. We are interested in the case
$1-d^2/4>m^2L^2>-d^2/4$. In this case, as first shown by
Breitenlohner and Freedman \brei, there are two ways to quantize
the field $\phi$ preserving the symmetries of AdS space. One can
impose the boundary condition $\alpha(\vec x)=0$, or the boundary
condition $\beta(\vec x)=0$.

 From a contemporary
point of view, as explained in \klebwit, this means that a
gravitational theory in AdS space that contains such a scalar
(along with other fields) has two different CFT duals on the
boundary, depending on which boundary condition one chooses to
impose.  If the boundary condition is $\alpha=0$, the boundary
theory has a conformal field ${\cal O}_\alpha$ of dimension
$\Delta_-$.  If one sets $\beta=0$, the boundary theory has a
conformal field ${\cal O}_\beta$ of dimension $\Delta_+$. Since
$\Delta_++\Delta_-=d$, we have $2\Delta_-<d$, so the $\alpha=0$
theory has a relevant operator ${\cal O}_\alpha^2$.  Perturbing
the $\alpha=0$ theory by this relevant operator, one gets a
renormalization group flow from the $\alpha=0$ theory to the
$\beta=0$ theory \ref\ugwitten{E. Witten, ``Multi-Trace Operators,
Boundary Conditions, and AdS/CFT Correspondence,''
hep-th/0112258.}. \nref\berkooz{M. Berkooz, A. Sever, and A.
Shomer, ``Double-trace Deformations, Boundary Conditions, and
Spacetime Singularities,''
JHEP {\bf 05} (2002) 034, hep-th/0112264.}%
(This flow is described by the more general boundary condition
$\alpha=f\beta$ \refs{\ugwitten,\berkooz}, where $f$ is the
coefficient of the relevant perturbation. See also \ref\minces{P.
Minces, ``Multi-trace Operators And The Generalized AdS/CFT
Prescription,'' hep-th/0201172} for more detail. Double-trace
perturbations in the AdS/CFT correspondence, such as ${\cal
O}_\alpha^2$, had been discussed earlier in \ref\silveratel{O.
Aharony, M. Berkooz, and E. Silverstein, ``Multiple Trace
Operators And Non-Local String Theories,'' JHEP {\bf 08} (2001)
006.}.  For more on the relation between conformal representations
corresponding to $\Delta_-$ and $\Delta_+$, see \ref\dob{V. K.
Dobrev, ``Intertwining Operator Realization Of The AdS/CFT
Correspondence,'' Nucl. Phys. {\bf B553} (1999) 559,
hep-th/9812194.}.)

\def\O{{\cal O}}
In the $\alpha=0$ theory, one would like to compute the generating
functional $\left\langle \exp(i\int d^3x\,J(\vec x)\O_\alpha(\vec
x))\right\rangle$ of correlation functions of $\O_\alpha$.  In the
AdS/CFT correspondence, this is done by computing the bulk
partition function with the boundary condition $\alpha(\vec x)=0$
generalized to $\alpha(\vec x)=J(\vec x)$ (where $J$ is a fixed
$c$-number source on the boundary of AdS space). Similarly, in the
$\beta=0$ theory, the generating functional
$\left\langle\exp(i\int d^3x\,J'(\vec x)\O_\beta(\vec
x))\right\rangle'$ of correlation functions of $\O_\beta$ is
computed by generalizing the boundary condition $\beta(\vec x)=0$
to $\beta(\vec x)= J'(\vec x)$, with $J'$ a fixed $c$-number
source.  (We use $\left\langle~~\right\rangle$ for the generating
function of unnormalized correlation functions in the $\alpha=0$
theory, and $\left\langle~~\right\rangle'$ for the analogous
function in the $\beta=0$ theory.)

It was found in \klebwit\ that taking the boundary of AdS to be
flat $\R^d$, and in the approximation of treating $\phi$ as a free
field, these two functionals are related by a Legendre
transformation. (Certain results in Liouville theory
\ref\klebold{I. R. Klebanov, ``Touching Random Surfaces And
Liouville Theory,'' Phys. Rev. {\bf D51}  (1995), 1836,
hep-th/9407167; I. R. Klebanov and A. Hashimoto,
``Non-perturbative Solution Of Matrix Models Modified By Trace
Squared Terms,'' Nucl. Phys. {\bf B434} (1995) 264,
hep-th/9409064.} gave a clue to this interpretation.) The Legendre
transformation is carried out as follows: one promotes the source
$J$ (or $J'$) of the $\alpha=0$ (or $\beta=0$) theory to a
dynamical field, couples it to a new source $J'$ (or $J$) via a
coupling $\int d^3x\,JJ'$, and performs the path integral over $J$
(or $J'$) plus  the other fields.

Though this relation for free field, flat space correlators puts
the $\alpha=0$ and $\beta=0$ theories on a completely symmetric
footing, in reality there is not that degree of symmetry between
them because there is a renormalization group flow from $\alpha=0$
to $\beta=0$ and not the other way around.  In more recent work
\nref\gubmit{S. Gubser and I. Mitra, ``Double-trace Operators And
One-Loop Vacuum Energy In AdS/CFT,'' hep-th/0210093.}%
\nref\gubkleb{S. Gubser and I. R. Klebanov, ``A Universal Result
On Central Charges In The Presence Of Double-Trace Deformations,''
hep-th/0212138.}%
\refs{\gubmit,\gubkleb}, the partition function of these theories
on ${\bf S}^d$, or equivalently their conformal anomaly, has been
investigated.  In \gubkleb, the following was demonstrated: the
partition function $Z_\beta$ of the infrared-stable theory with
boundary condition $\beta=0$ can be obtained from the path
integral of the $\alpha=0$ by letting the source $J$ of the
$\alpha=0$ theory become dynamical and integrating over it along
with the other fields: \eqn\turcy{Z_\beta=\int DJ \left\langle
\exp\left(i\int d^3 x \,J{\cal O}_\alpha\right)\right\rangle.}
This is an analog for the partition function in curved space of
the statement about flat space correlation functions made in the
last paragraph. (We could combine the two statements by adding a
new source $J'$ for the operator $\O_\beta$ of the $\beta=0$
theory, with a coupling $\int d^3x\,JJ'$.)  This more refined
statement does not have a counterpart with $\alpha$ and $\beta$
exchanged.

We want to give an alternative explanation of the result \turcy.
Then, going back to gauge theory, we will offer a similar
explanation for the relation between the $S$ operations in bulk
and on the boundary. We place an infrared cutoff on the theory by
truncating AdS space near the conformal boundary.  So henceforth,
instead of $X$ denoting the full AdS space with $Y$ as conformal
boundary, $X$ is a compact manifold and $Y$ is its ordinary
boundary. We want to compare the partition function $Z_{\rm free}$
of a scalar field $\phi$ on $X$ with free boundary conditions --
in which the boundary values of $\phi$ are unrestricted -- to a
path integral with Dirichlet boundary conditions, in which $\phi$
vanishes on the boundary.  (Free boundary conditions are also
called Neumann boundary conditions, as they lead by the equations
of motion to vanishing of the normal derivative of $\phi$.)

To make this comparison, we write an arbitrary field $\phi$ on $X$
as \eqn\arbfi{\phi=\phi_0+\tilde\phi,} where $\phi_0$ vanishes on
the boundary and $\tilde\phi$ is any function on $X$ that agrees
with $\phi$ on the boundary. Since we will be integrating over
$\phi_0$ (and a change in $\tilde\phi$ can be absorbed in a shift
in $\phi_0$) it does not matter exactly how $\tilde\phi$ is
chosen.  If $\phi$ is treated as a free field, $\tilde\phi$ can
conveniently be chosen as the unique solution of the classical
equations of motion that coincides with $\phi$ on the boundary.

If we simply set $\phi=\phi_0$ with $\tilde \phi$ absent, the path
integral over $\phi$ gives the partition function $Z_{\rm Dir}$ of
the theory with Dirichlet boundary conditions.  Instead of simply
setting $\tilde\phi$ to zero, let us specify it and hold it fixed.
In performing the path integral over $\phi_0$ with fixed
$\tilde\phi$,  we regard $\tilde\phi$ as the source for an
operator $\O_{\rm Dir}$ in the Dirichlet theory. So the path
integral over $\phi_0$ with fixed $\tilde \phi$ computes the
unnormalized generating functional $\left\langle\exp\left(i\int
d^3x\,\tilde\phi \,\O_{\rm Dir}\right)\right\rangle$.  If now we
integrate over $\tilde\phi$, the combined path integral over
$\phi_0$ and $\tilde\phi$ is the same as the path integral over
$\phi$, and should give $Z_{\rm free}$. So we expect
\eqn\narbfi{Z_{\rm free}=\int
D\tilde\phi\left\langle\exp\left(i\int d^3x\,\tilde\phi \,\O_{\rm
Dir}\right)\right\rangle.}

This relation has an obvious analogy with the result \turcy\ of
\gubkleb.  The reason for the relation seems clear intuitively.
Going back to \noffo, we have $z^{\Delta_-}>>z^{\Delta_+}$ for $z$
small.  So a path integral with $\beta=0$ corresponds in the
theory that has a cutoff at very small $z$ to a path integral with
$\phi$ vanishing on the boundary, that is, with Dirichlet boundary
conditions. And a path integral with $\alpha=0$ corresponds to a
path integral with Neumann or free boundary conditions, the
boundary value of $\phi$ being unrestricted.

\bigskip\noindent{\it The $S$ Operation}

Now let us return to our problem of understanding the relation
between the $S$ operation for abelian gauge fields in
four-dimensional AdS space and the $S$ operation in the boundary
conformal field theory.

We consider a $U(1)$ gauge field $A$ on a cutoff version of AdS
space -- a compact (but large) manifold $X$ with boundary $Y$.
$\vec B=0$ boundary conditions are the analogs for gauge fields of
Dirichlet boundary conditions for scalars -- they say that $A$
vanishes on the boundary, up to a gauge transformation.  $\vec
E=0$ boundary conditions are analogous to free or Neumann boundary
conditions.  They leave the boundary values of $A$  unrestricted.
The analog of the above argument says that the path integral of
the $\vec E=0$ theory is obtained by adding the boundary value of
$A$ as an additional field and integrating over it as well as over
the other variables, which include the choice of $A$ in the
interior. This indicates that electric-magnetic duality in bulk
gives the $S$ operation of the boundary theory.

\bigskip\noindent{\it Partial Generalization For Nonabelian Gauge
Theory}

Consider an AdS theory which contains, instead of the $U(1)$ gauge
field that we have considered, a nonabelian gauge field with
unbroken gauge group $G$. In this case, unless special collections
of matter fields are included, we do not have an $SL(2,{\bf Z})$
duality of the low energy gauge theory on AdS, so we will not get
an $SL(2,{\bf Z})$ action on possible dual conformal field
theories. Nonetheless, a few of the things we have said do
apparently generalize to the nonabelian case via the same
arguments that we have given above.

Nonabelian gauge theory in the bulk can be quantized with (at
least) the two possible conformally invariant boundary conditions
$\vec B=0$ and $\vec E=0$.  So this will give two possible dual
CFT's.  The $\vec B=0$ theory has $G$ as a global symmetry,
generated by an adjoint-valued conserved current $J$. The $\vec
E=0$ theory is obtained from the $\vec B=0$ theory by coupling a
gauge field $A$ (without kinetic energy) to $J$.

Furthermore, one can consider the operation $\theta\to\theta+2\pi$
in the bulk theory.  Applied to the $\vec B=0$ theory, this
operation merely shifts the two-point function of $J$ by a contact
term.  Applied to the $\vec E=0$ theory, it shifts the
Chern-Simons level of the gauge field $A$.

\bigskip
This work was supported in part by NSF Grant PHY-0070928.
\listrefs
\end